\documentstyle[twoside,fleqn,espcrc2]{article}

\newcommand{\AmS}{{\protect\the\textfont2
  A\kern-.1667em\lower.5ex\hbox{M}\kern-.125emS}}

\title{Perturbative QCD and Tau Decay\thanks{Invited talk 
presented at the TAU 96 Workshop, Estes Park, Colorado, September 1996.}
}%

\author{Eric Braaten\address{Department of Physics, The Ohio State
University, Columbus, OH  43210, U.S.A.}%
}

\begin{document}


\begin{abstract}
Sufficiently inclusive observables in the decay of the tau lepton
can be calculated using the methods of perturbative QCD.
These include the asymmetry parameter 
$A_\tau$ that determines that angular distribution of the total hadron 
momentum in the decay of a polarized tau.  It should be possible to 
measure $A_\tau$ accurately using existing data from LEP.
Reliable estimates of theoretical errors are essential in order to 
determine whether a given observable is sufficiently inclusive to be
calculated using perturbative methods.  The theoretical uncertainties 
due to higher orders in $\alpha_s$ can be estimated using recent 
calculations to all orders in the large--$(33-2N_f)$ limit.  These 
estimates indicate that tau decay data can be used to determine 
$\alpha_s(M_Z)$ to a precision of 2\% or better.
\end{abstract}

\maketitle

\section{INTRODUCTION}

What can perturbative QCD tell us about the decays of the tau lepton?  If
you ask this question to the typical man on the street, he will answer
``Absolutely nothing!''  Perturbative QCD tells us about the interactions
of quarks and gluons with large momentum transfer.  But the decay of the
$\tau$ is dominated by decays into single particles and resonances:
$\pi$, $\rho$, $a_1,$ etc.  The QCD interactions that bind quarks and
gluons into these hadrons necessarily involve small momentum transfers and
are therefore completely outside the domain of perturbation theory.
Therefore one should not expect perturbative QCD to tell us
anything about the decays of the $\tau$.

But suppose we ignore this objection, and proceed to calculate the decay
rate of the tau into a neutrino plus quarks and gluons using
perturbative QCD.  The resulting expression for the hadronic decay rate is
an expansion in powers of $\alpha_s$.  Using the measured value of the
hadronic decay rate, we can determine $\alpha_s$.  We find that this value
for the QCD coupling constant agrees amazingly well with the best
determinations of $\alpha_s$.  Is this just a remarkable coincidence?  Or
is it possible that sufficiently inclusive observables in tau decay
really can be calculated using the methods of perturbative QCD?

\section{QCD PREDICTION FOR $R_\tau$}

The QCD prediction for the ratio $R_\tau$ of the hadronic and electronic
branching fractions of the tau has the form \cite{B-N-P}
\begin{eqnarray}
R_\tau &=& 3 \;(|V_{ud}|^2 + |V_{us}|^2)\; S_{EW} \cr
&&\times \{ 1 + \delta_{EW} + \delta_{pert} + \delta_{power}\}.
\end{eqnarray}
The electroweak corrections include a multiplicative factor $\delta_{EW} =
1.0194$ and a small additive correction $\delta_{EW}=
0.0010$ \cite{Marciano-Sirlin}.  There is a perturbative QCD corrrection
$\delta_{pert}$ that can be expressed as an expansion in powers of
$\alpha_s$ at the scale $M_\tau$ (or any other scale $\mu$ of your
choosing) \cite{Schilcher-Tran}:
\begin{eqnarray}
\delta_{pert} &= &{\alpha_s(M_\tau) \over \pi} + 5.2 \left({\alpha_s(M_\tau)
\over \pi} \right)^2  \cr
&& + \, 26.4   \left({\alpha_s(M_\tau)
\over \pi} \right)^3 + \;\dots .
\end{eqnarray}
The remaining QCD corrections are the power corrections $\delta_{power}$,
the most important of which are the following \cite{Narison-Pich}:
\begin{eqnarray}
\delta_{power} &= & -16 {\overline{m^2} \over
M_\tau^2} +\, 32 \pi^2 \, {\langle m \overline \psi \psi  \rangle
\over M_\tau^4}\cr
&&+ \,{11 \pi^2 \over 4} \left ( {\alpha_s(M_\tau) \over \pi}
\right )^2 {\langle {\alpha_s \over \pi} GG\rangle \over M^4_\tau} \cr
&&   + \mbox{ terms of the form } {\langle \bar \psi \Gamma
\psi \bar \psi \Gamma \psi \rangle \over M_\tau^6},
\end{eqnarray}
where $\overline {m^2}$ is a weighted average of the running quark
masses $m_u^2$, $m_d^2$, and $m_s^2$ evaluated at the scale $M_\tau$ with
weights $1/2$, $|V_{ud}|^2/2$, and $|V_{us}|^2/2$, respectively.
Similarly, $\langle m \overline \psi \psi \rangle$ is the same weighted
average of the quark condensate matrix elements $\langle m_u \overline u u
\rangle$, $\langle m_d \overline d d \rangle$, and $\langle m_s \overline s
s \rangle$.  The matrix element $\langle {\alpha_s \over \pi} G G
\rangle$ is called the gluon condensate.

The QCD prediction (1) is based on the fact that the inclusive hadronic
decay rate of the tau involves only 2 momentum scales:  the mass
$M_\tau$ and $\Lambda_{QCD}$, the scale associated with
nonperturbative effects in QCD.  One can imagine increasing $M_\tau$ while
holding the QCD coupling constant, and therefore  $\Lambda_{QCD}$, fixed.
It is possible to systematically separate the effects of ``hard'' partons,
whose momenta scale with $M_\tau$, from the effects of ``soft'' partons,
whose momenta remain proportional to  $\Lambda_{QCD}$ as $M_\tau$
increases.  At very short distances of order $1/M_W$, the decay of the
$\tau^-$ proceeds through the decay into a neutrino plus $d \bar u$ (or
$s \bar u$).  The QCD corrections involve both hard partons and soft
partons.  The perturbative correction $\delta_{pert}$ consists of
corrections from the emission of hard partons and from the exchange of
hard partons.  It can be calculated as a perturbation series in
$\alpha_s(M_\tau)$.  The first 3 terms in this series are known and they
are given in (2).

What about the effects of soft partons? There are large corrections from
the emission of soft gluons and large corrections from the exchange of
soft gluons, but they cancel order-by-order in $\alpha_s$.  In other
words, the dominant effects of the soft gluons can be expressed as a
unitary transformation that does not change the decay rate.  At the
perturbative level, this unitary transformation describes the evolution of
an initial state consisting of hard partons emerging from the decay into
final states consisting of hard partons and soft partons.  At the
nonperturbative level, the soft gluons have dramatic effects, binding the
hard partons into final state hadrons.  Nevertheless, the dominant effect
of the soft gluons can still be expressed as a unitary transformation
which evolves an initial state consisting of hard partons into final
states consisting of hadrons.  Since a unitary transformation preserves
probabilities, it does not change the decay rate.

Now the effects of the soft partons are not exactly a unitary
transformation.  There are corrections which can be taken into account
through power corrections $\delta_{power}$, like those given in (3).  
The term  involving the
gluon condensate $\langle {\alpha_s \over \pi} GG\rangle$ takes into
account effects of soft gluons that can not be expressed as a unitary
transformation.  The quark condensate $\langle m \bar \psi \psi \rangle$
takes into account corrections from soft quarks and
antiquarks that are suppressed by the quark mass.  
The matrix elements of the form $\langle \bar \psi \Gamma
\psi \bar \psi \Gamma \psi \rangle$ take into account corrections from 
soft quarks and antiquarks that would be present even if the
quarks were massless.

The QCD predictions for all inclusive observables in tau decay have the
same general form as the predictions for $R_\tau$ in (1).  There is a
free-quark prediction that corresponds to the decay into a neutrino plus $d
\bar u$ or $s \bar u$.  There are perturbative corrections that can
be expressed as an expansion in $\alpha_s(M_\tau)$ like that in (2).
Finally, there are power corrections like those given in (3).  
They include corrections from running quark masses and soft
parton corrections that are expressed in terms of vacuum matrix elements.

\section{OBSERVABLES}

Below, I enumerate the observables in tau decay that can be calculated
using the methods of perturbative QCD.

\subsection{Inclusive decay rate}

The QCD prediction for the inclusive hadronic decay rate normalized to the
electronic decay is given in (1).  The ratio $R_\tau$ can be further
resolved into the contributions from the weak vector current and the weak
axial-vector current.  It can also be resolved into ``non-strange''
contributions proportional to $|V_{ud}|^2$ and ``strange'' contributions
proportional to $|V_{us}|^2$ \cite{B-N-P}.

\subsection{Invariant-mass distribution}

Let $s$ be the square of the invariant mass of the hadrons in the decay of
the $\tau$.  The distribution $dR_\tau/ds$, or more accurately the
moments of this distribution, can be calculated using perturbative
methods.  The $0^{th}$ moment is simply $R_\tau$ itself.  The QCD
predictions for the moments are given in Ref. \cite {deLiberder-Pich}.
These moments have been used to measure the matrix elements that appear in
the power corrections (3) \cite {Duflot}.

\subsection{Angular distribution}

There is one inclusive observable in tau decay that has not yet been
measured experimentally.  This is a parameter $A_\tau$ that determines
the angular distribution of the total hadron momentum in the decay of a 
polarized tau.  Let $\theta$ be the angle in the tau rest frame
between the total hadron momentum and the quantization axis for the spin
of the $\tau$.  If the tau is unpolarized, the angular distribution is
uniform in $\cos \theta$.  If the tau has polarization $P$, the angular
distribution is
\begin{equation}
{d R_\tau \over d \cos \theta} = {1 \over 2} R_\tau (1 + A_\tau P \cos
\theta).
\end{equation}
The size of the asymmetry in $\cos \theta$ is determined by the parameter 
$A_\tau$.  This parameter is known for various exclusive decay modes.  It
has the value +1 if the hadronic final state is a single $\pi$.  
The value is close to
$-1$ for the transverse polarization modes of the $\rho$ and $a_1$, and close
to +1 for the longitudinal polarization mode.  The inclusive asymmetry
parameter $A_\tau$ is the weighted average over all hadronic final states
of the asymmetry parameter for each of the exclusive decay modes.

The asymmetry parameter $A_\tau$ can be calculated using the mothods of
perturbative QCD with the same degree of rigor as $R_\tau$ itself
\cite{Braaten}.  The free-quark prediction for
$A_\tau$ is ${1 \over 3}$.  The QCD corrections increase the prediction to
$A_\tau = 0.415 \pm 0.022$.  Thus the contributions of the $\pi$, $\rho$,
$a_1$ and all the multihadronic modes must conspire to give a value that is
about 25\% larger than the free quark value.

An ``experimental'' value for $A_\tau$ has recently been obtained by
Weinstein \cite {Weinstein}.  He generated 50,000 polarized tau decay
events by running the Monte Carlo programs KORALB and TAUOLA and obtained
the value $A_\tau = 0.385 \pm 0.007$.  These programs use the known
branching fractions for the various hadronic decay modes.  They incorporate
the correct asymmetry parameters for the $\pi$, $\rho$, and $a_1$ modes,
but they are not tuned to give the correct asymmetries for the multihadronic
modes.  Thus a real experimental measurement of $A_\tau$ is needed in order
to test the QCD prediction.

A good way to measure $A_\tau$ is to use the taus from $Z^0$ decay,
because they are naturally polarized with polarization $P = -0.14$.
Suppose one could assemble an unbiased sample of taus from $Z^0$
decay.  If, for each tau decay, one measured the invariant mass $s$ 
of the hadrons and their total energy $E$ in the $Z^0$ rest frame, then
$A_\tau$ could be determined by calculating the following average over the
$\tau$ sample:
\begin{equation}
\left \langle {1-2E/M_Z \over 1-s/M_\tau^2}\right \rangle = {1 \over 2}
\left(1- {1 \over 3} A_\tau P \right).
\end{equation}
From simple statistics, one would need a sample of at least 2000 taus
to distinguish the free quark prediction 0.508 for (5), which corresponds to
$A_\tau = {1 \over 3}$, from the unpolarized value of 0.5.  One
would need at least 30,000 taus to distinguish the QCD prediction
$0.5097 \pm 0.0005$ from the free quark prediction.  With a sample
of 400,000 taus from $Z^0$ decay, it should be possible to measure
$A_\tau$ rather accurately.  Note that the prediction
that the asymmetry parameter for all the exclusive modes must average out
to a value near the free quark value
${1 \over 3}$ is already rather remarkable and thus even a
crude measurement of $A_\tau$ would be useful.  It would be the first 
confirmation of a perturbative QCD prediction for  spin-dependent 
observables in tau decay.  An accurate measurement of
$A_\tau$ in agreement with the QCD prediction would provide dramatic
evidence that inclusive observables can  
be accurately calculated using the methods of perturbative
QCD.  It should lay to rest any questions of whether the accuracy of the
value of $\alpha_s$ obtained from $\tau$ decay is merely fortuitous.

\section{ERROR ESTIMATES}

\begin{table*}[hbt]
\setlength{\tabcolsep}{1.5pc}
\newlength{\digitwidth} \settowidth{\digitwidth}{\rm 0}
\catcode`?=\active \def?{\kern\digitwidth}
\caption{Estimates of the theoretical errors (in units of $10^{-4}$)
in the value of $\alpha_s(M_Z)$ determined from tau decay.}
\label{tab:Estimates of Errors}
\begin{tabular*}{\textwidth}{@{}l@{\extracolsep{\fill}}lrr}
\hline
& \qquad Altarelli & Narison \\ \cline{2-3} \\
running:  $M_\tau \to M_z$ \qquad \qquad  &\qquad  ~~~20 & 10~~~ \\
freezing of $\alpha_s$                    &\qquad  ~~~10 & --~~~ \\
quark masses                              &\qquad ~~~~-- & 5~~~ \\
condensates                               &\qquad ~~~~-- & 9~~~ \\
higher orders in $\alpha_s$               &\qquad  ~~~65 & 14~~~ \\
$\mu$ dependence                          &\qquad ~~~~-- &  9~~~ \\
renormalization scheme                    &\qquad ~~~~-- &  5~~~ \\
``other theoretical errors''              &\qquad  ~~~10 & --~~~ \\
\hline
Total Error                               &\qquad  ~~~70~ & 23~~~ \\
\hline
\end{tabular*}
\end{table*}

In order to make quantitative tests of the QCD predictions for the
observables described in the last section, it is essential to have
reliable estimates of the theoretical errors.  In the proceedings of
TAU 94, both Altarelli and Narison made attempts to estimate the
theoretical errors in the determination of $\alpha_s(M_Z)$ from $\tau$
decay.  Their estimates are given in Table 1.  The bottom line is that
Altarelli's estimate of the theoretical error is larger than Narison's by
a factor of 3.  This factor of 3 makes all the difference in the world.
If Narison's estimate is correct, then tau decay remains competitive
with the best determinations of $\alpha_s$ that are available.  If
Altarelli's estimate is correct, then the determination of $\alpha_s$ from
tau decay is only interesting as a qualitative test of perturbative QCD.

Before assessing the error estimates of Altarelli and Narison, it is
necessary to discuss some of the theoretical issues that are involved in
these estimates.

\subsection{Renormalization scale $\mu$}

The ratio $R_\tau$, if calculated to all orders in $\alpha_s$, would be
independent of the renormalization scale $\mu$.  However, if the
perturbation series is truncated at some order in $\alpha_s$, it will depend
on $\mu$.  The value of $\alpha_s$ determined by measuring
$R_\tau$ will therefore depend on the choice for $\mu$.

This can be illustrated by considering the perturbation series truncated
after the order$-\alpha_s$ term.  If we choose the renormalization scale
to be $\mu = M_\tau$, the prediction for $R_\tau$ (ignoring electroweak
and power corrections and setting $|V_{ud}|^2 + |V_{us}|^2 = 1$ for
simplicity) is
\begin{equation}
R_\tau = 3\left\{ 1 + {\alpha_s(M_\tau) \over \pi}\right\}.
\end{equation}
If we choose the scale to be some fraction $x$ of the tau
mass $\mu = xM_\tau$, then the prediction is
\begin{equation}
R_\tau = 3\left\{ 1 + {\alpha_s(x M_\tau) \over \pi}\right\}.
\end{equation}
Having determined $\alpha_s(x M_\tau)$ from the measured value of
$R_\tau$, we can then determine $\alpha_s(M_\tau)$ by using the
renormalization group:
\begin{equation}
\alpha_s(M_\tau) = {\alpha_s (xM_\tau) \over 1+ {9 \over 2 \pi} \alpha_s
(xM_\tau) \log (1/x)}.
\end{equation}
The resulting numerical value of $\alpha_s(M_\tau)$ will differ from that
obtained directly from (6).  The difference
between (6) and the combination of (7) and (8) amounts to summing certain
terms of the form $\alpha_s^n \log^n(M_\tau/\mu)$ to all orders in $n$.

The dependence on the renormalization scale $\mu$ decreases if the
perturbation series is calculated to higher order in $\alpha_s$.  However,
as long as the perturbation series is truncated, there will always be some
dependence on $\mu$.

\subsection{Truncation of the perturbation series}

In (2), we have truncated the perturbation expansion for $R_\tau$
in powers of $\alpha_s(M_\tau)$, but this is not the only
way to truncate the perturbation series.  The ratio
$R_\tau$ can be expressed in the form of a contour integral
\begin{eqnarray}
R_\tau &=& 12 \pi^2 {1 \over 2 \pi i} \oint_{|t|=M_\tau^2} {dt \over t} 
\cr  
&& \hspace {-1cm} 
\times \left (1-2 {t \over M_\tau^2} + 2{t^3
\over M_\tau^6} - {t^4\over M_\tau^8} \right)D\big(\alpha_s (-t) \big),
\end{eqnarray}
where the function $D(\alpha_s)$ has a perturbation expansion in powers of
$\alpha_s$:
\begin{eqnarray}
D(\alpha_s) &=& {1 \over 4 \pi^2} \Big\{ 1 \;+\; {\alpha_s \over \pi}  
\;+\; 1.6 \left({\alpha_s \over \pi}\right)^2 \cr
&& \hspace {1cm} 
\;+\; 6.4 \left({\alpha_s \over \pi}\right)^3 \Big\}.
\end{eqnarray}
The contour integral in (9) runs counterclockwise around the circle $|t| =
M_\tau^2$, beginning at $t=M_\tau^2 + i \epsilon$ and ending at
$t=M_\tau^2- i \epsilon$.  The function $\alpha_s (-t)$ in (9) is
the analytic continuation of the running coupling constant to the 
complex $t$-plane.  It is defined by the boundary condition 
$\alpha_s(-t) = \alpha_s(M_\tau)$  at $t=-M_\tau^2$ 
and by the differential equation 
$t (d / dt)\alpha_s (-t) = \beta \big( \alpha_s(-t) \big)$, 
where $\beta (\alpha_s)$
is the beta function for QCD:  $\beta (\alpha_s) = - (9 / 4 \pi)
\alpha_s^2 + \dots$.

One way to truncate the perturbation series is to first carry out the
contour integral and then truncate the expansion for $R_\tau$.  This gives
the usual perturbation series for $R_\tau$ given in (2).  An alternative
possibility is to truncate the expansion for $D(\alpha_s)$ and then carry
out the contour integral analytically.  This has been advocated in
particular by deLiberder and Pich
\cite{deLiberder-Pich}.  The resulting expression for $R_\tau$ is
\begin{eqnarray}
R_\tau &=& 3 \Bigg\{ 1 \;+\; \overline{{\alpha_s (-t) \over \pi}} \;+\; 1.6
\overline{\left( {\alpha_s (-t) \over \pi} \right )^2}\cr
&& \hspace {1cm} 
\;+\; 6.4 \overline{\left ({\alpha_s (-t) \over \pi}\right )^3}\, \Bigg\}\,,
\end{eqnarray}
where the coupling constant with the bar over it represents the following
weighted average of its values along the contour in the complex plane:
\begin{eqnarray}
\overline{\alpha_s^n (-t)} &=& {1 \over 2 \pi} \int_0^{2 \pi} d \theta
\;(1 -  2e^{i \theta} + 2 e^{3 i \theta} - e^{4 i \theta})\cr
&&  \hspace {2cm} 
\times \alpha_s^n(-M_\tau^2 e^{i \theta}).
\end{eqnarray}

The effect of this prescription for calculating $R_\tau$ is more easily
illustrated using the simpler example of the ratio $R_{e^+ e^-}$.  It can
also be expressed as a contour integral involving the same function
$D(\alpha_s)$ as in (9):
\begin{equation}
R_{e^+e^-} ( \sqrt {s}) = 12 \pi^2 {1 \over 2 \pi i} \oint_{|t|=s} {dt
\over t} D\big( \alpha_s (-t) \big).
\end{equation}
We choose the center-of-mass energy $\sqrt s$ of the $e^+ e^-$ to be below
charm threshold, so that the number of light quark flavors is 3, just as
in tau decay.  If we integrate and then truncate $R_{e^+ e^-}$, then at
one loop, we obtain
\begin{equation}
R_{e^+e^-} (\sqrt s) = 3 \left\{1+{\alpha_s (\sqrt s) \over \pi}\right \}.
\end{equation}
If we truncate $D(\alpha_s)$ and then integrate, we obtain
\begin{equation}
R_{e^+e^-} (\sqrt s) =  3 \left \{1+\overline{{\alpha_s (-t) \over \pi}}
\right \},
\end{equation}
where $\overline{\alpha_s (-t)}$ is $\alpha_s(-se^{i \theta})$ averaged
over the angles $\theta$.  Expressed in terms of $\alpha_s(\sqrt s)$, this
average is ${4 \over 9} \arctan \big( {9 \over 4} \alpha_s(\sqrt s)
\big)$.  Its expansion in powers of $\alpha_s(\sqrt s)$ is
\begin{eqnarray}
\overline{\alpha_s (-t)} &=&
\alpha_s (\sqrt s) \Big\{ 1 \;-\; {27 \pi^2 \over 16} \big(
{\alpha_s ( \sqrt s) \over \pi} \big)^2 \cr
&&
\;+\; {6561 \pi^4 \over 1280} \left(
{\alpha_s ( \sqrt s) \over \pi} \right)^4 +
\dots \Big \}.
\end{eqnarray}
Thus the expression (15) for $R_{e^+ e^-}$ differs from
(14) by the resummation of terms of the form $\pi^{2n}
(\alpha_s/\pi)^{2n}$ to all orders in $n$.

Similarly, the perturbative expansion (11) for $R_\tau$ differs 
from the one given in (2) by
the resummation of terms of the form $\pi^{2n} (\alpha_s / \pi)^{2n}$ to all
orders in $n$.  DeLiberder and Pich give a number of arguments why this
resummation should be preferred, but I don't find them
convincing.  One argument is that it greatly decreases the dependence of
$R_\tau$ on the renormalization scale $\mu$.  This is true, but there is
an equally convincing argument against the resummation, and that is that
the power series expansion for $D(\alpha_s)$ is more severely divergent
than that for $R(\alpha_s)$.

\subsection{Divergence of the perturbation series}

It has been known for a long time that most perturbation expansions in
field theory, including QED and QCD, are actually divergent series.
Suppose you were able to calculate the correction of order $\alpha_s^n$
for any value of $n$.  For small values of $n$, you might find that
adding more and more correction terms gives a better and better
approximation.  But if you continued adding higher and higher orders in
$\alpha_s$, you would eventually find that the approximation would
get worse and worse.  The order in $\alpha_s$ at which the series
begins to diverge depends on the value of $\alpha_s$, decreasing roughly
as $1/\alpha_s$.  Thus the divergence of the perturbation series is a
much more important issue for tau decay than it is for applications of
perturbative QCD at higher energy, where the running couping constant is
smaller.

There has been a dramatic development in the last five years that has
made the problem of the divergence of the perturbation series much more
concrete.  A subset of the Feynman diagrams for $R_\tau$ and $R_{e^+e^-}$
have been calculated to all orders in $\alpha_s$ \cite{Beneke}.  The
diagrams are those of order $\alpha_s^n$ with the maximum number
$n-1$ of quark loops.  The complete perturbation series for $R_\tau$ has the
form
\begin{equation}
R_\tau = 3 \left\{ 1 + {\alpha_s (M_\tau) \over \pi} + \sum_{n=2}^\infty r_n
\alpha_s^n (M_\tau)\right\}.
\end{equation}
The coefficients $r_n$ are polynomials in the number $N_f$ of light quark
flavors, or equivalently, in $33-2N_f$:
\begin{eqnarray}
r_n &=& r_n^{(0)} \;+\; r_n^{(1)} (33-2N_f) \;+\; \dots \cr 
&& \hspace{1cm} \
\;+\; r_n^{(n-1)} (33-2N_f)^{n-1}.
\end{eqnarray}
It is the numbers $r_n^{(n-1)}$ that have been computed to all orders in
$n$.

Given the coefficients $r_n^{(n-1)}$ defined in (18), we can calculate a
subset of the higher order corrections to $R_\tau$ to all orders in
$\alpha_s$.  I will denote the resulting expression by $\hat R_\tau$ and
refer to it as the large -- $(33-2N_f)$ limit:
\begin{equation}
\hat R_\tau \equiv 3 \left\{ 1 + {\alpha_s(M_\tau) \over \pi} +
\sum_{n=2}^\infty \hat r_n \alpha_s^n (M_\tau) \right\},
\end{equation}
where $\hat r_n = r_n^{(n-1)} (33-2N_f)^{n-1}$.  The first few terms in
the expansion for $\hat R_\tau$ are
\begin{eqnarray}
\hat R_\tau &=& 3 \; \bigg\{ 1 \;+\; {\alpha_s (M_\tau) \over \pi} 
\;+\; 5.1 \left({\alpha_s (M_\tau) \over \pi}\right)^2 \cr
&& \hspace{-.5cm} \;+\; 28.8 \left({\alpha_s(M_\tau) \over \pi} \right)^3 
\;+\; 156.7 \left({\alpha_s (M_\tau) \over \pi} \right)^4 \cr
&& \hspace{-.5cm} \;+\; 900.8 \left({\alpha_s (M_\tau) \over \pi}\right )^5 
\;+\; \dots \bigg\}.
\end{eqnarray}
This expansion can be compared with the exact expansion for the
corrections to $R_\tau$, which is given in (2).  Note the remarkable
agreement between the coefficients of $\alpha_s^2$ and $\alpha_s^3$ in the
two expansions.  This gives us some confidence that the large--$(33-2N_f)$
limit can predict correctly the sign and order of magnitude of the
coefficient of the $\alpha_s^4$ term.

Since we know the expansion (19) for $R_\tau$ in the large--$(33-2N_f)$
limit to all orders in $\alpha_s$, we can use it to study the behavior of
the perturbation expansion \cite{Maxwell}.  One finds that 
(19) is a divergent series with coefficients $\hat r_n$ that
grow asymptotically like $n!$:
\begin{equation} \hat r_n \to {2 \over 15 \pi} e^{-53} \left(- {9 \over 4
\pi} \right)^n n! \;.
\end{equation}
Thus the sum in (19) if taken literally is meaningless.  However, there is
a standard procedure for recovering an analytic function from the
divergent power series generated by its Taylor expansion. It is called
{\it Borel resummation}.  From the series (19) for $\hat R_\tau$, one
constructs the Borel transform $\hat B(b)$ by dividing each coefficient by
$(n-1)!$:
\begin{equation}
\hat B(b) = \sum_{n=1}^\infty {\hat r_n \over (n-1)!} b^{n-1}.
\end{equation}
The power series for $\hat B(b)$ is  less divergent than that for
$R_\tau$, and  actually converges for 
$|b| < 4 \pi / 9$.

An analytic expression for $\hat B(b)$ is known.  It has poles at integer
multiples of $4 \pi / 9$.  There are poles on the negative real axis at
$b= n(4 \pi / 9)$, $n= -1, -2, \dots$ that are called {\it ultraviolet
renormalons}.  There are also poles on the positive real axis at $b=n(4
\pi / 9)$, $n=2, 3, \dots$ that are called {\it infrared renormalons}.
Given the analytic expression for the function $\hat B(b)$ defined by
(22), we can recover the desired function $\hat R_\tau$ by computing the
inverse Borel transform:
\begin{equation}
\hat R_\tau = 3\left\{ 1 + \int_0^\infty db \, \exp\left ({-b \over \alpha_s
(M_\tau)}\right) \, \hat B(b) \right\}.
\end{equation}
The infrared renormalon poles on the integration contour can be handled
using a principal value prescription.

Now the divergence of the perturbation series for $\hat R_\tau$ is
dominated by the singularity of $\hat B(b)$ that is closest to the
origin.  This is the first ultraviolet renormalon, which is a pole at $b=
-4 \pi / 9$.  The behavior of $\hat B(b)$ near this pole is
\begin{equation}
\hat B(b) \longrightarrow {2 \over 15 \pi} e^{-5/3} {1 \over 1 + {9 \over 4
\pi} b}.
\end{equation}
The remainder is analytic inside the circle $|b| < 8 \pi / 9$.  The
pole term can be expanded as a power series in $b$, but the series has a
radius of convergence of only $4 \pi /9$.  Thus perturbation theory gives
a convergent approximation to the function $\hat B(b)$ in the integrand of
(23) only in the interval $0 < b < 4 \pi /9$.  Outside that interval, the
pole term given in (24) cannot be approximated by a truncated perturbation
series.  The integral of the pole term from $4 \pi / 9$ to infinity
therefore gives a ``renormalon correction'' to $\hat R_\tau /3$:
\begin{eqnarray}
\delta_{ren} &=& {2 \over 15 \pi} e^{-5/3} \int_{4 \pi /9}^\infty
db \; \exp \left(-{b \over\alpha_s (M_\tau)}\right)\cr
&& \hspace{2cm} \times {1\over 1 + {9 \over 4 \pi} b}.
\end{eqnarray}
The asymptotic value of the integral for small values of $\alpha_s
(M_\tau)$ is
\begin{equation}
\delta_{ren} = {1 \over 15} e^{-5/3} {\alpha_s(M_\tau) \over
\pi} \exp \left(-{4 \pi \over 9 \alpha_s(M_\tau)}\right).
\end{equation}
Finally, using the expression  $2\pi / \big( 9 \log \, (M_\tau / \Lambda)
\big)$ for the running coupling constant $\alpha_s(M_\tau)$, the leading
renormalon correction reduces to
\begin{equation}
\delta_{ren} = {1 \over 15} e^{-5/3} {\alpha_s(M_\tau) \over
\pi} {\Lambda^2 \over M_\tau^2},
\end{equation}
where $\Lambda$ is the renormalization group invariant scale parameter in
the $\overline{MS}$ scheme.

The analysis above was carried out for the standard truncation of the
perturbation series for $R_\tau$.  With deLiberder-Pich resummation, one
must apply Borel resummation to the function $D(\alpha_s)$ in the
integrand of (9).  The Borel transform of this function has a double pole
at $b = -4 \pi / 9$.  The renormalon correction analogous to (25) therefore
has two terms corresponding to the double pole and a single pole.  
After integrating over $t$, it reduces to
\begin{eqnarray}
\delta_{ren} &=& {2 \over 9 \pi} e^{-5/3} \int^\infty_{4\pi/9} db
\exp \left ( - {b \over \alpha_s (M_\tau)} \right) \cr
&& \times 
\left [ {2 \over (1 + {9 \over 4 \pi} b )^2} 
	+ {5 \over 1 + {9 \over 4 \pi} b} \right ] \cr
&& \times 
{12 \sin ( {9 \over 4} b) \over 
{9 \over 4} b (1 - {9\over 4 \pi} b ) ( 3 - {9 \over 4 \pi} b) 
	(4- {9 \over 4 \pi} b) } .
\end{eqnarray}
The asymptotic value of the integral for small values of $\alpha_s
(M_\tau)$ is
\begin{equation}
\delta_{ren} = {4 \over 3} e^{-5/3} {\alpha_s (M_\tau) \over
\pi} {\Lambda^2 \over M_\tau^2}.
\end{equation}
Note that this is larger by a factor of 20 than the 
corresponding correction (27) for
the standard truncation.  This simply reflects the fact that the
perturbation series for $D(\alpha_s)$ diverges more severely than that for
$R_\tau$.

\subsection{Estimates of the perturbative error}

We are now in a position to assess the error estimates of Altarelli and
Narison.  The perturbation series for $R_\tau$ can be written
\begin{eqnarray}
R_\tau &=& 3 \bigg\{ 1 \;+\; \sum_{n=1}^3 r_n \alpha_s^n(M_\tau) 
	\;+\; r_4 \alpha_s^4 (M_\tau) \cr
&& \hspace{1cm} 
\;+\; \sum_{n=5}^\infty r_n \alpha_s^n (M_\tau) \bigg\}.
\end{eqnarray}
The first 3 correction terms in the expansion are given in (2).  We have
separated the unknown $\alpha_s^4$ term from the sum of all the
higher-order corrections.

Narison's error estimate is based on the assumption that the error from
the $\alpha_s^4$ term in (30) dominates over that from the sum of all
higher orders.  He used the values of the lower-order coefficients $r_1$,
$r_2$, and $r_3$ to guess a reasonable range for the values of $r_4$.  The
resulting estimate of the error from higher orders is
\begin{equation}
(\Delta_{pert})_{\rm Narison} = \pm 50 \left({\alpha_s (M_\tau)
\over
\pi}\right)^4.
\end{equation}
Translating this into an error on $\alpha_s(M_Z)$, we obtain the value
$\pm 14 \times 10^{-4}$ given in Table 1.

\begin{table*}[hbt]
\setlength{\tabcolsep}{1.5pc}
\catcode`?=\active \def?{\kern\digitwidth}
\caption{Estimates of the theoretical error in $\alpha_s(M_\tau)$ (in units
of
$10^{-4}$) from higher orders in $\alpha_s(M_\tau)$.}
\label{tab:Estimates of Theoretical Errors}
\begin{tabular*}{\textwidth}{@{}l@{\extracolsep{\fill}}lrr}
\hline
& \qquad error from & \quad error from sum \\
& \qquad $n=4$ term & \quad of $n \ge 5$ terms\\ \cline{2-3} \\
large--$(33-2N_f)$ limit (truncation in $\alpha_s(M_\tau)$) &\qquad\qquad 40
& 0.2~~~~~~~ \\
large--$(33-2N_f)$ limit (deLiberder-Pich resummation) &\qquad \qquad ~4 &
4~~~~~~~ \\
Altarelli & \qquad\qquad ~-- & 65~~~~~~~ \\
Narison & \qquad\qquad  14 & --~~~~~~~ \\
\hline
\multicolumn{3}{@{}p{120mm}}{}
\end{tabular*}
\end{table*}

Altarelli's estimate was based on the assumption that the error from the
sum of all higher order terms in (30) dominates over that from the
$\alpha_s^4$ term.  As shown in the previous subsection, there is a
renormalon contribution to that sum that is proportional to $\Lambda^2
/ M_\tau^2$.  Altarelli's error estimate is
\begin{equation}
(\Delta_{pert})_{\rm Altarelli} = \pm {1 \over 4 } {
\Lambda^2
\over M_\tau^2}.
\end{equation}
Translating this into an error on $\alpha_s(M_Z)$, we obtain the value
$\pm 65 \times 10^{-4}$ given in Table 1.  There is no apparent
calculation underlying the coefficient $1/4$ in (32).  It seems to be pure
guesswork on the part of Altarelli.

One of the problems with both of these error estimates is that they
require guessing the magnitude of unknown coefficients:  $r_4$ in the case
of Narison, the coefficient of $\Lambda^2 /
M_\tau^2$ from the renormalon correction in the case of Altarelli.  One way
to avoid such guesswork is to use the explicit calculations to all orders
in the large--$(33-2N_f)$ to estimate the errors.  As an estimate of the
error from the $\alpha_s^4$ term in (30), we can use the magnitude of the
$\alpha_s^4$ term in the large--$(33-2N_f)$ limit:
\begin{equation}
\Delta_{pert} = \pm | \hat r_4 | \alpha_s^4
(M_\tau).
\end{equation}
As an estimate of the error from the sum of all higher order terms in
(30), we can use the magnitude of the sum of all higher order terms in the
large--$(33-2N_f)$ limit:
\begin{equation}
\Delta_{pert} = \pm
\left |
\sum_{n=5}^\infty \hat r_n \alpha_s^n (M_\tau) \right |.
\end{equation}
The divergent series on the right side of (34) is defined by Borel
resummation.  A simpler estimate, which can be 
compared directly with that of Altarelli, is to pick out the leading renormalon
correction, which is given in (27) for the standard truncation in
$\alpha_s(M_\tau)$ and in (29) for deLiberder-Pich resummation.

The error estimates based on the large--$(33-2N_f)$ limit 
are given in Table 2. The standard
truncation gives a much larger error from the $\alpha_s^4$ term and a much
smaller error from the sum of all higher-order terms.  With
deLiberder-Pich resummation, there is a better balance between these two
sources of uncertainty.  Adding the two errors, we take the total error in
$\alpha_s(M_Z)$ from higher orders in $\alpha_s$ to be $8 \times
10^{-4}$.  The error estimates of Altarelli and Narison are also given in
Table 2 for comparison.  The error estimate from the large--$(33-2N_f)$
limit is consistent with that of Narison, but much smaller than that of
Altarelli.  The explicit calculations in the large--$(33-2N_f)$ limit
indicate that the coefficient of the $\Lambda^2 /
M_\tau^2$ correction arising from the first ultraviolet renormalon 
is an order of magnitude smaller than assumed by Altarelli.

\section{CONCLUSIONS}

Sufficiently inclusive observables in tau decay can be calculated using
the methods of perturbative QCD.  A particularly interesting observable
that has yet to be measured is the asymmetry parameter $A_\tau$ that
describes the angular distribution of the total hadron momentum in the
decay of a polarized tau.  Using existing data from LEP, it should be
possible to measure this parameter with sufficient precision to
discriminate between the free-quark prediction of ${1 \over 3}$ and the QCD
prediction $A_\tau = 0.41 \pm 0.02$.  Such a measurement would be 
the first test of QCD predictions for spin-dependent observables
in tau decay and it would
demonstrate conclusively that methods based on perturbative QCD give
accurate predictions for sufficiently inclusive observables.

In order to determine whether a given observable in tau decay is
sufficiently inclusive to calculate using perturbative methods, it is
essential to have reliable estimates of the theoretical errors.  One can
use recent calculations to all orders in $\alpha_s$ in the
large--$(33-2N_f)$ limit to estimate the errors from the higher-order terms
in the perturbation series.  The resulting error estimate on
$\alpha_s(M_Z)$ is consistent with the 1994 estimate of Narison,
but much smaller than the error estimate of Altarelli.  These explicit
calculations indicate that Altarelli
overestimated by an order of magnitude the size of $\Lambda^2/M_\tau^2$
corrections associated with the first ultraviolet renormalon.  
The bottom line is that
$\alpha_s(M_Z)$ can indeed be determined from $\tau$-decay data with a
precision of 2\% (or perhaps even better).

This work 
was supported in part by the U.S.
Department of Energy, Division of High Energy Physics, under 
Grant DE-FG02-91-ER40684.

\end{document}